\pdfoutput=1
\documentclass[aps,prl,reprint,amsmath,amssymb,nofootinbib,superscriptaddress]{revtex4-1}
\usepackage{custom}
\begin{document}

\title{Sign optimization and complex saddle points in one-dimensional QCD}

\author{G\"{o}k\c{c}e Ba\c{s}ar}
\email{gbasar@unc.edu}
\author{Joseph Marincel}
\email{marincel@unc.edu}
\affiliation{Department of Physics and Astronomy, University of North Carolina, Chapel Hill, North Carolina 27599, USA}

\date{\today}

\begin{abstract}   
We study one-dimensional QCD at finite quark density by using the sign optimization framework. The fermion sign problem is mitigated by deforming the path integral domain, $SU(3)$ to a complexified one $\M\subset SL(3)$, explicitly constructed to reduce the phase fluctuations. The complexification is constructed using the angular representation of $SU(3)$. We provide a physical explanation of the optimization procedure in terms of complex saddle points. This picture connects the sign optimization framework to the generalized Lefschetz thimbles.  

\end{abstract}

\maketitle

\section{Introduction}
\label{sec:intro}

Understanding the phases of matter governed by the strong nuclear force remains an open problem, even after fifty years since the formulation of the microscopic theory, Quantum Chromo Dynamics (QCD). The main difficulty in comprehensively mapping the phase diagram of QCD quantitatively is the strong interactions between quarks and gluons which necessitates non-perturbative methods. A standard, non-perturbative approach is lattice QCD which heuristically formulates QCD as a classical statistical system such that physical observables can be calculated numerically via importance sampling. The probability of each configuration is given by $e^{-S}/Z$ where S is the (Euclidean) action and $Z$ is the partition function. However, for a large class of problems, including QCD with finite quark density, $S$ is complex, and the importance sampling suffers from large phase oscillations \cite{deForcrandreview,Philipsen:2007aa,KARSCH200014,Aarts_2016}. This is known as the \textit{sign problem}. In fact, the sign problem is ubiquitous in systems with strongly interacting fermions at finite density, such as the Hubbard model away from half-filling, which is thought to model high $T_c$ superconductors \cite{PhysRevB.40.506}, or neutron matter at the core of neutron stars, as well as out-of-equilibirum systems that evolve in real-time where the Feynman path integral is by construction complex \cite{Berges:2006xc,Alexandru:2016gsd,Mou:2019tck,Kanwar:2021tkd,Lawrence:2021izu}. 

Recently a set of frameworks has been developed to tackle the sign problem which utilize the fact that one could change the domain of the field theory path integral to a complexified one without changing the value of the integral. One then finds a domain which reduces the phase oscillations thereby reducing the sign problem, which is similar to but more general than the multi-dimensional stationary phase contours (i.e. Lefschetz thimbles) \cite{Cristoforetti:2012su,Fujii:2013sra,Alexandru:2015xva,DiRenzo:2020vou}. Variations on this idea include generalized Lefschetz thimbles \cite{Alexandru:2015sua}, sign optimization \cite{Mori:2017pne, Alexandru:2018fqp,Kashiwa:2019lkv}, ``Learnifolds" \cite{Alexandru:2017czx} to name a few. See also \cite{Alexandru:2020wrj} for a recent review of these ideas and an extensive set of references. 

In this work we explore the sign optimization approach in one-dimensional QCD where the complex domain is built explicitly by minimizing the sign problem using an optimization method. A similar analysis of one-dimensional QCD has been done in Ref.~\cite{Mori:2019tux}. One shortcoming of these kinds of optimization / Machine Learning type approaches is that what happens between the input and output is generally not visible, making it challenging to have picture of the process. Because of this ``black-box" nature of the method, it is often difficult to make use of the specific properties (e.g. symmetries) of the underlying theory which for complicated systems such as QCD is likely necessary.  In this work, we focus on providing a physical picture of the optimization procedure in terms of complex saddle points of the underlying theory. Our work also provides a natural connection between the optimization and Lefschetz thimble approaches. 

\section{One Dimensional QCD}
\label{sec:QCD01}
In this section we quickly review the essential properties of one-dimensional QCD, \cite{Bilic:1988rw} which has been studied to test various ideas and formalisms to tackle QCD related problems, such as the properties of the Dirac spectrum \cite{Ravagli:2007rw}, and the sign problem \cite{Aarts:2010gr, Bloch:2013ara}, in particular by using a similar sign optimization method to ours \cite{Mori:2019tux} and  Lefschetz thimbles \cite{DiRenzo:2017igr}. 

We shall consider an $SU(3)$ gauge theory with $N_f$ number of quarks on a lattice with no spatial extent and $N_t$ sites in the temporal direction which sets the temperature $T=1/N_t a$ where $a$ is the lattice spacing. We assume all the quarks have mass $m$ and consider the system at finite density with the associated chemical potential $\mu$.  
In one dimension, there is no plaquette (field strength) hence no gauge action. After integrating out the fermions, the only remaining degrees of freedom are the gauge links, $U_t$, that wrap around the temporal direction. Furthermore we can fix the gauge such that only the last link  is non-trivial, i.e $P\equiv U_1\dots U_{N_t}$. In short, the theory is essentially a matrix model of Polyakov loops, $P\in SU(3)$, with the partition function
\bea
Z= \int dP\, {\det}^{N_f}(aD)\,
\label{eq:pf}
\ea 
where $dP$ is the Haar measure and the Dirac determinant is given as
\bea
\label{eq:Dirac_op}\small
\det(aD)=\frac1{2^{3N_t}}\det\paranth{A{\mathbb 1}_{3 \times 3} + e^{\mu/T}P + e^{-\mu/T}P^\dagger}\,\,\quad
\ea
with $A=2\cosh(\mu_c/T)$ and $\mu_c=a^{-1}\sinh(am)$. At finite density, the theory exhibits the sign problem as $\det(aD)$ becomes complex for nonzero values of $\mu$. The sign problem essentially kicks in around $\mu\approx\mu_c$ which approaches $m$ (the lowest energy to excite a state) in the continuum limit. A standard way of dealing with the sign problem is ``reweighting" where the field configurations are sampled from the probability distribution $e^{-\Re S}$. The remaining phase, $ e^{-i\Im S}$, is treated as a part of the observable. The physical observables can be expressed in terms of the ``phase quenched" theory with the partition function $Z_{pq}=\int dP e^{-\Re S}$ as
\begin{equation}
\av{{\cal O}}\equiv\frac1Z{\int dP e^{-S}\,{\cal O} }= \frac1\sigma\frac1Z_{pq}{\int dPe^{-\Re S}\, {\cal O}e^{-i\Im S} }\quad
\end{equation}
where the average phase,
\begin{equation}
\sigma=\av{e^{-i\Im S}}_{\Re S}\equiv\frac{\int dPe^{-\Re S}\,e^{-i\Im S}}{\int dPe^{-\Re S}}=\frac{Z}{Z_{pq}}\leq1,
\label{eq:sigma}
\end{equation}
can be viewed as a measure for the severity of the sign problem. Small values of $\sigma$ indicate a small overlap between the original theory and the phase quenched one which means one needs to sample a large number of configurations (typically exponentially) in the phase quenched theory in order to obtain an accurate value for the observable. This is a manifestation of the sign problem. 

Finally, it is useful to parameterize $SU(3)$ in terms of 8 angles, $\Phi_i\in\{\theta_1,\theta_2,\theta_3,\phi_1,\dots,\phi_5\}$, 
with $0\leq\theta_i\leq \pi/2$ and $0\leq\phi_i\leq2\pi$ \cite{Bronzan:1988wa} such that the Polyakov loop can be written as
\begin{widetext}
\begin{equation}\label{eq:Bronzan}
P=
\begin{pmatrix}
c_1 c_2 e^{i \phi_1} & s_1 e^{i \phi_3} & c_1 s_2 e^{i \phi_4} \\
s_2 s_3 e^{-i(\phi_4+\phi_5)}-s_1 c_2 c_3 e^{i(\phi_1+\phi_2-\phi_3)} & c_1 c_3 e^{i \phi_2} & -c_2 s_3 e^{-i(\phi_1+\phi_5)} - s_1 s_2 c_3 e^{i(\phi_2-\phi_3+\phi_4)} \\
-s_1c_2s_3 e^{i(\phi_1-\phi_3+\phi_5)}-s_2c_3 e^{-i(\phi_2+\phi_4)} & c_1s_3 e^{i\phi_5} & c_2c_3 e^{-i(\phi_1+\phi_2)}-s_1s_2s_3 e^{i(-\phi_3+\phi_4+\phi_5)} \\
\end{pmatrix}
\end{equation}
\end{widetext}
where $c_i\equiv \cos\theta_i$, and $s_i=\sin\theta_i$ and the Haar measure is
\begin{equation}
dP = H({\vec\theta}) d^8\Phi,\quad H({\vec\theta})=\frac{1}{2\pi^5} s_1 c_1^3 s_2 c_2 s_3 c_3 \,.
\end{equation}

\section{Complexification and sign optimization}
\label{sec:optimization}

The domain of the path integral \eqref{eq:pf} is $SU(3)$. However by Cauchy's theorem we can deform it into another domain $\M \subset  SL(3)$. As long as $\M$ is continuously connected to $SU(3)$ without crossing any singularities, and the integrand is a holomorphic function, as we shall assume, the value of the path integral does not change. At the same time, since $S_R$ is \textit{not} a holomorphic function, the denominator in Eq.~\eqref{eq:sigma}, $Z_{pq}$ \textit{does} change. The key idea is to find a surface $\M$ over which $Z_{pq}$ is smaller, leading to a larger value of $\sigma$ and a milder sign problem. 

We will do this by using the sign optimization method where we parameterize $\M$ by using a set of variables, $\vl$, and maximize $\sigma$ with respect to $\vl$. We first parameterize the complex space $\M$ as
\begin{equation}
\tilde\Phi_i=\Phi_i + i f_{\Phi_i}(\Phi)
\end{equation}
Here $ f_{\Phi_i}(\Phi)=0$ corresponds to $SU(3)$. The eight complex angles $\tilde\Phi_i$ parameterize an eight-dimensional surface in $SL(3)$ which can be continuously connected to $SU(3)$ via a family of intermediate surfaces defined through $f\rightarrow s f$ with $0\leq s\leq1$. An advantage of using this parameterization is that the path integral over $\M$ can easily be re-expressed in terms of the original variables as:
\bea
Z=\int d^8\Phi J(\Phi)e^{-S_{H}(\tilde\Phi(\Phi))}\equiv \int d^8\Phi e^{-S_{e}(\Phi)}
\ea
where $J(\Phi)=|\det(\del \tilde\Phi_i /\del \Phi_j)|$ is the associated Jacobian for the change of variables from $\tilde\Phi$ back to $\Phi$, $S_{H}(\Phi)\equiv-N_f\log\det(aD)-\log H(\Phi)$ and $S_e(\Phi)\equiv S_{H}(\tilde\Phi(\Phi))-\log J(\Phi)$. Each function $f_{\Phi_i}$ can in principle depend on all eight angles but must obey the periodicity conditions $f_{\theta_i}(\theta_i=0) = f_{\theta_i}(\theta_i=\pi/2)=0$, $f_{\phi_i}(\phi_i=0) = f_{\phi_i}(\phi_i=2 \pi)$ \cite{Detmold:2021ulb,Kanwar:2021wzm}. In this work we will use two different ansatze for $f_{\Phi_i}s$: (1) the ``diagonal" ansatz:\footnote{In what follows we will simply refer to the Fourier coefficients, $\lambda^{(i)}\cos\eta_i,\lambda^{(i)}\sin\eta_i $ and $\nu^{(i)}$s collectively as $\vec \lambda$.}
\bea
\label{eq:diagonal}
 f_{\phi_i}=\sum_{n=0}^{N-1} \lambda^{(i)}_{n}\cos(n\phi_i+\eta^{(i)}_{n}), f_{\theta_i}=\sum_{n=1}^{N-1} \nu^{(i)}_{n}\sin(2n\theta_i)
\quad
\ea
and (2) the ``mixing" ansatz:
\bea
\label{eq:mixing}
f_{\phi_i}=\begin{cases}\sum_{m,n=0}^{N-1} \lambda^{(i)}_{m,n}\cos(m\phi_1+n\phi_2+\eta^{(i)}_{mn}),
\,\, {i=1,2}
\\
\lambda^{(i)},  \,\, {i=3,4,5}
\end{cases}
\ea
Working with these more restricted anstaze rather than the most general Fourier decomposition reduces the number of parameters in the optimization procedure making it practically feasible. We elaborate on this further below. 

The next step is to find a value of $\vl$ that maximizes $|\sigma|$. To achieve this we follow a gradient ascent algorithm to find a local extremum of $\log|\sigma_{\vl}|$\footnote{Here we explicitly denote the dependence of $\sigma$ on $\M$ with the subscript $\vl$}. We first start from $SU(3)$ (i.e. $\vl=0$) and update $\vl$ according to 
\begin{equation}
\label{eq:grad_ascent}
\vl(\tau+1) = \vl(\tau)+ \delta \nabla_{\vl}\log |\sigma_{\vl(\tau)}|  
\end{equation}
where $\tau$ enumerates the gradient ascent steps. We repeat this procedure up to some $\tau_{max}$. Depending on the physical parameters such as $\mu, N_f$, etc., the step size $\delta$ has to be empirically adjusted so that it is neither too large, which leads to runaways, nor too small, which leads to slow convergence to minimum. More sophisticated adaptive algorithms can also be used if needed. A straightforward calculation leads to 
\bea
\label{eq:dsigma}
\nabla_{\vl}\log |\sigma_{\vl}| &=&  \av{\Re(\nabla_{\vl} S_e -\textup{Tr} (J^{-1} \nabla_{\vl} J))}_{S_R}\qquad
\ea
where $S_R\equiv \Re S_H$ \cite{Alexandru:2018ddf}. Therefore each gradient ascent update requires a Monte-Carlo computation which remarkably does not have any sign problem. In the following section we present the results of this procedure.

\section{Results}
\label{sec:results}

\begin{figure}
\includegraphics[scale=0.45]{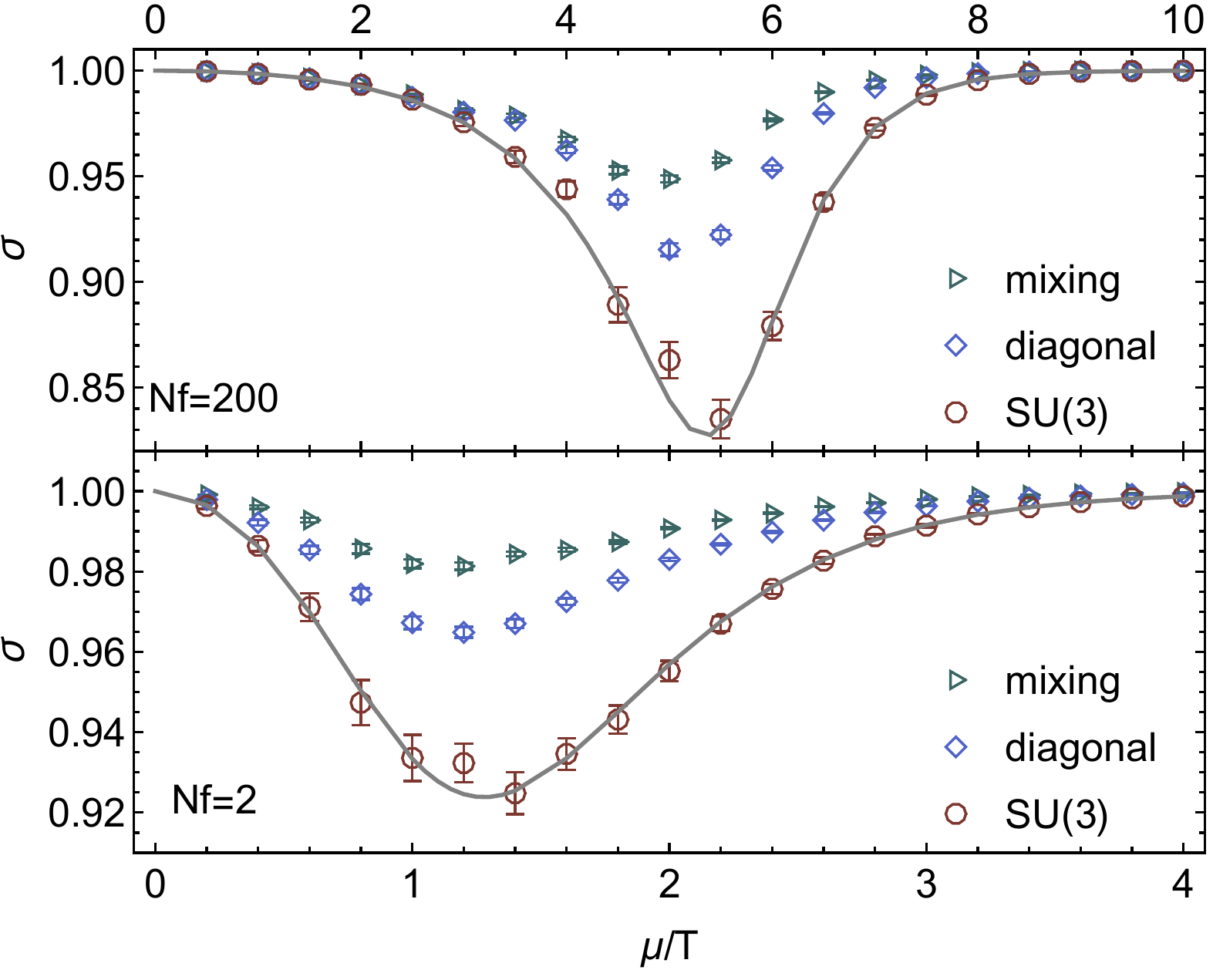}
\caption{The average sign as a function of chemical potential for $N_f=2,200$. Solid curves represent the exact value of $\sigma$ for $SU(3)$. }
\label{fig:sign}
\end{figure}

\begin{figure}
\includegraphics[scale=0.41]{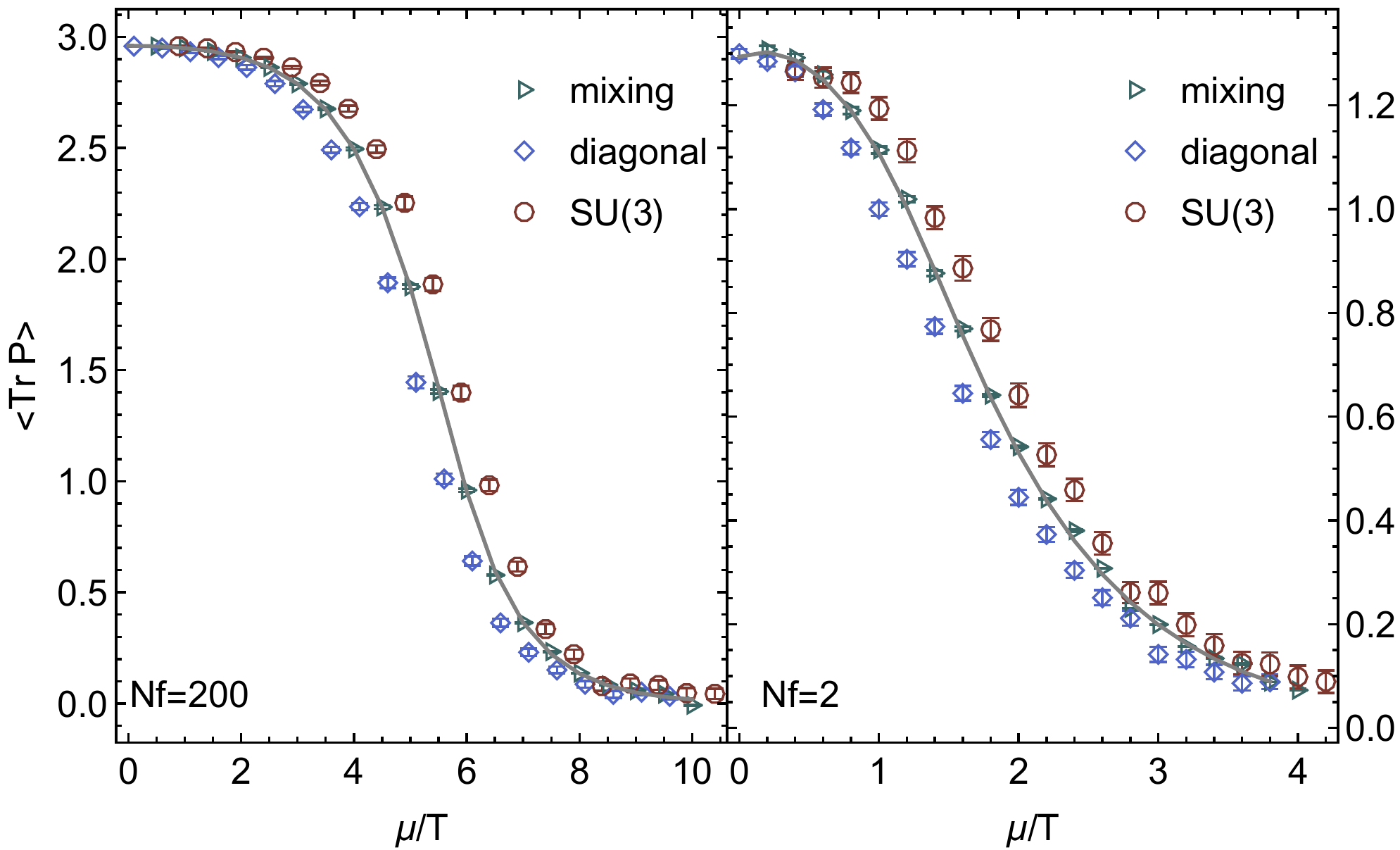}
\caption{The Polyakov loop, $\av{ {\rm Tr}  P}$ as a function of $\mu/T$. For visual clarity the $SU(3)$ and diagonal ansatz data are offset in the $x$-axis. }
\label{fig:polyakov}
\end{figure}

We performed sign optimization for 2 and (to compare with semi-classical estimates) 200 quark flavors, and for a range of $\mu$. We worked in the chiral limit, $m=0$, since it is the limit where the sign problem kicks in as quickly as possible (i.e. $\mu_c=0)$. We set the lattice spacing $a=1$ and all the dimensionful quantities are measured in units of temperature. In sign optimization we have used $N=4$ Fourier coefficients for the diagonal (44 parameters) and $N=3$ for the mixing (39 parameters) ansatze.  Each Monte-Carlo run had $5000$ configurations in each gradient ascent step generated by a standard Metropolis algorithm where the remaining sign problem is reweighted. The observables in Figs.~\ref{fig:sign} and \ref{fig:polyakov} calculated on $\M_{\vl}$ are obtained from the last step of the gradient ascent. 

\begin{figure}
\includegraphics[scale=0.39]{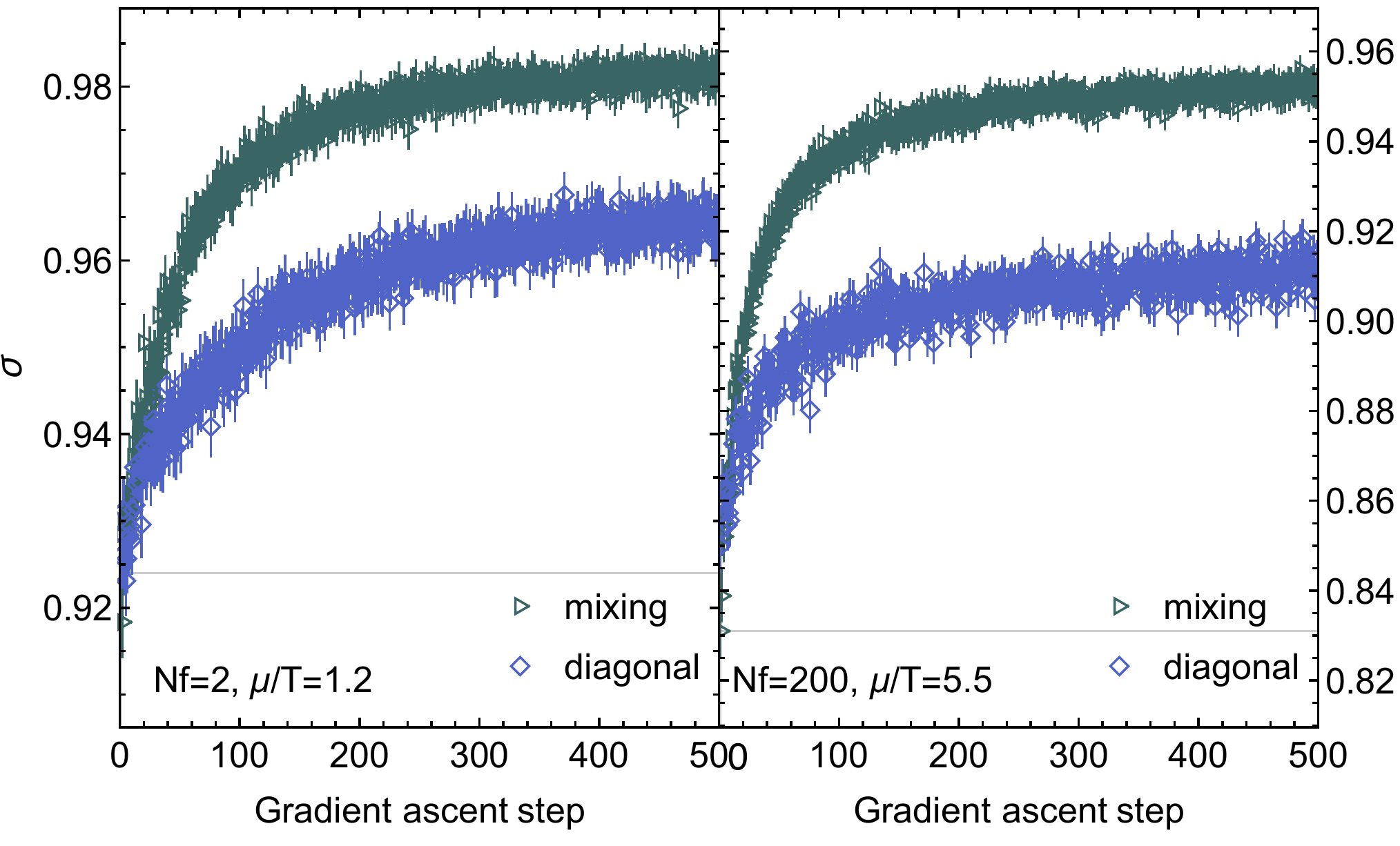}
\caption{The improvement of the sign problem as measured by the average phase, $\sigma$, through the gradient ascent. Solid lines represent the exact value of $\sigma$ for $SU(3)$. }
\label{fig:ga}
\end{figure}

In Fig.~\ref{fig:sign} we show the average phase, $\sigma$, as a function of $\mu$ compared with the exact analytical result for $SU(3)$. For any $N_f$ and $\mu$ the mixed ansatz defined in performs better than the diagonal. As expected the uncertainty in the results decreases as the sign improves. In Fig.~\ref{fig:ga} we show the improvement in the sign problem (as gauged by $\sigma$) as a function of the gradient ascent step for values of $\mu$ where the original sign problem is the worst (see Fig.~\ref{fig:sign}).

Finally in Fig.~\ref{fig:polyakov} we show the average Polyakov loop as a function of $\mu$ compared with the exact analytical result. As expected all the results agree with the exact result, a consequence of the fact that the $\M_{\vl}$ is equivalent to $SU(3)$ as a path integral domain albeit with a milder sign problem. Even though the original sign problem is not too severe to begin with, the improvement in the sign problem can be seen in the noticeably smaller error bars, especially for the mixing ansatz. 

\section{Discussion and Semiclassical Analysis}
\label{sec:semiclassical}

\begin{figure}
\includegraphics[scale=0.42]{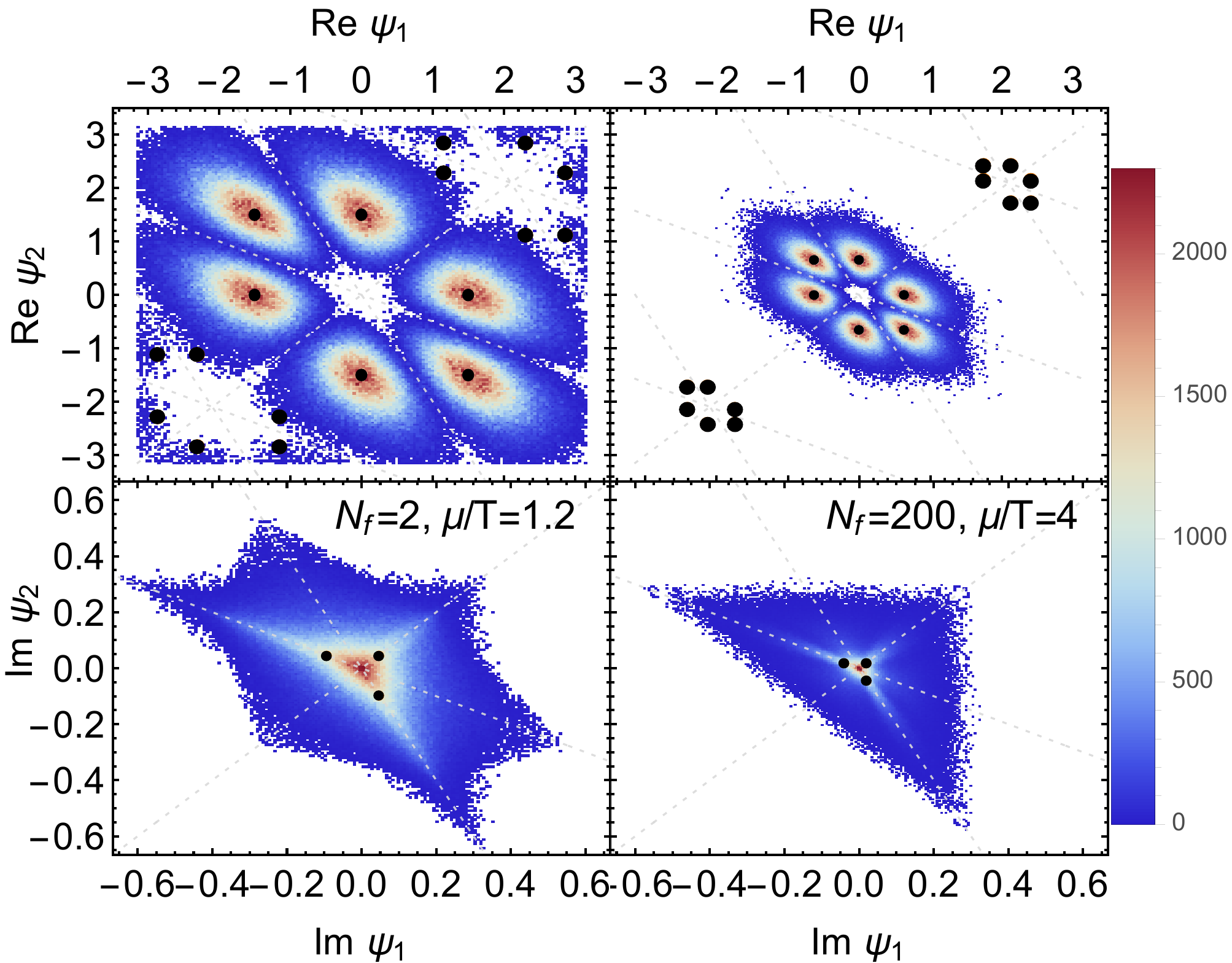}
\caption{The distribution of the eigenvalues of the Polyakov loop. The locations of the complex saddle points are denoted by black dots. The dashed lines are $y=x,-2x,-x/2$.}
\label{fig:hist}
\end{figure}

Let us now discuss the results and present a physical picture for the sign optimization procedure. In general it is difficult to visualize eight dimensional surfaces, however it is useful to work with the eigenvalues of the Polyakov loop. For any $P$ we can write $U P U^\dagger={\rm diag}(e^{i\psi_1},e^{i\psi_2},e^{i\psi_3})$ such that $\psi_1+\psi_2+\psi_3=0$ for some $U\in SU(3)$. Now, the path integral can be expressed in terms of two independent eigenvalues, say $\psi_{1,2}$. The Haar measure reduces to $dP\propto V(\psi_1,\psi_2) d\psi_1d\psi_2$ where 
$
V(\psi_1,\psi_2)=\sin^2(\frac{\psi_1-\psi_2}{2})\sin^2(\frac{2\psi_1+\psi_2}{2})\sin^2(\frac{\psi_1+2\psi_2}{2})
$
is a Vandermonde determinant. In Fig.~\ref{fig:hist} we show the distribution of the eigenvalues for two sets of parameters $N_f=2$, $\mu=1.2$ and $N_f=200$, $\mu=4$ generated by using the mixing ansatz. The real parts of $\psi_{1,2}$ are distributed around six regions related to Weyl chambers of $SU(3)$, whereas the imaginary parts (associated with the deviation from $SU(3)$) are mostly concentrated approximately in a triangular area. For $N_f=200$ the configurations are clustered closer to $\psi_i=0$.

The physics of this pattern can be understood as follows. The saddle points of the path integral are given by $\del_{\psi_i} S_{eff}(\psi_1,,\psi_2)=0$ where $S_{eff}=-N_f \log\det(aD)-\log V$. They are complex valued as a result of the interplay between the Dirac determinant which attracts the eigenvalues towards one of the three center elements of $SU(3)$\footnote{The three center elements of $SU(3)$ are $\psi^{(0)}_1=\psi^{(0)}_2=0,\psi^{(1)}_1=\psi^{(1)}_2=2\pi/3$, and $\psi^{(2)}_1=\psi^{(2)}_2=4\pi/3$.}, and the Haar measure which is repulsive. In the semiclassical, large $N_f$ limit the eigenvalues approach\footnote{However, the measure of the center values in the path integral is always zero for any finite $N_f$.} to the center values, where the imaginary part vanishes. In other words the Haar measure splits each naive eigenvalue, $\psi^{(k)}_i$ into six clusters of complex eigenvalues distributed around it, one per each chamber (see Fig.~\ref{fig:hist}). Each eigenvalue in a given cluster has the same action. Furthermore for any eigenvalue, $\psi_i$, $-\psi^*_i$ is also an eigenvalue due to the underlying ${\cal CK}$ symmetry \cite{Nishimura:2014rxa}. Similar phenomena is observed in other matrix models \cite{Buividovich:2015oju} 

From Fig.~\ref{fig:hist} one can see that the sign optimization ``builds" the complex domain, $\M_{\vl}$, around the complex saddles, within the constraints of the ansatz. This picture also sheds light on why the mixing ansatz outperforms the diagonal one even though it only depends on two of the eight angles and depends on smaller number of parameters. In a suitable gauge, $P$ can be expressed as  ${\rm diag}(e^{i\phi_1},e^{i\phi_2},e^{-i(\phi_1+\phi_2)})$. Therefore having terms such as $\cos(n_1\phi_1+n_2\phi_2+\eta)$ in the ansatz allows it to capture the fluctuations around the complex saddle points (i.e. the Lefschetz thimbles) more accurately than the diagonal ansatz.

\section{Conclusions}
\label{sec:conclusions}

In this work we studied the sign problem in one-dimensional QCD. The main strategy we followed was to construct complex path integration domains that reduce the sign problem without changing the value of the path integral. These domains are constructed explicitly by maximizing the average sign that is a gauge for the sign problem via gradient ascent. We observed that the sign optimization constructs the domains around the complex saddle points of the theory even though it has no prior knowledge of them. Roughly speaking, it approximately re-constructs the generalized thimbles around these saddles points. This can be seen more clearly in the semiclassical limit, but persists even away from it.  We also showed that an ansatz for the optimization that allows more freedom to explore the fluctuations around the saddles performs better.  The main takeaway is that the knowledge of complex critical points can be used to tailor the optimization ansatz to maximize performance, highlighting the importance of studying complex saddles in quantum field theory \cite{Nishimura:2014rxa,Behtash:2015loa}.

\acknowledgments
\section{Acknowledgments}

GB is supported by the National Science Foundation CAREER Award PHY-2143149. GB thanks Gerald Dunne for helpful comments.

\bibliographystyle{utphys}
\bibliography{refs}

\end{document}